\documentclass[11pt]{amsart} 
\usepackage[latin9]{inputenc}
\usepackage{geometry}
\usepackage[english]{babel}
\usepackage{mathrsfs}
\usepackage{amsmath}
\usepackage{amssymb}
\usepackage{esint}
\usepackage[unicode=true,pdfusetitle,
 bookmarks=true,bookmarksnumbered=false,bookmarksopen=false,
 breaklinks=false,pdfborder={0 0 1},backref=false,colorlinks=false]
 {hyperref}

\newtheorem{Lemma}{Lemma}[section]

\newtheorem{Thm}[Lemma]{Theorem}

\newcommand{\BigOh}[1]{\mathcal{O}(#1)}

\begin{document}

\title[The wave equation  
       in Friedmann -- Robertson -- Walker space-times]
     {On the initial value problem for the wave equation \\ in
      Friedmann -- Robertson -- Walker space-times}
\author{Bilal Abbasi \& Walter Craig$^*$}
\address{Department of Mathematics \& Statistics \\
         McMaster University \\
         Hamilton, Ontario L8S~4K1, {\sc Canada} \\ 
         and $^*$The Fields Institute \\ 222 College Street, 
         Toronto, Ontario M5T 3J1 {\sc Canada}}

\begin{abstract}
The propagator $W(t_0,t_1)(g,h)$ for the wave equation in a given space-time
takes initial data $(g(x), h(x))$ on a Cauchy surface $\{(t,x)\, : \, t=t_0 \}$ 
and evaluates the solution $(u(t_1,x),\partial_t u(t_1,x))$ at other times
$t_1$. The Friedmann -- Robertson -- Walker space-times are defined
for $t_0, t_1 > 0$, while for $t_0 \to 0$ there is a metric
singularity. Klainerman and Sarnak~\cite{Klainerman&Sarnak} give a
spherical means representation for the general solution of the wave
equation with the Friedmann -- Robertson -- Walker background metric
in the three spatial dimensional cases of curvature $K = 0$ 
and $K = -1$. We derive from the expression of their representation
three results about the wave  propagator for the Cauchy problem in these 
space-times.  Firstly, we give an elementary proof of the sharp rate of time
decay of solutions with compactly supported data. Secondly, we observe
that the sharp Huygens principle is not satisfied by solutions, unlike
in the case of three-dimensional Minkowski space-time (the usual
Huygens principle of finite propagation speed is satisfied, of
course). Thirdly we show that for $0 < t_0 < t$ the limit   
\[
     \lim_{t_0 \to 0+} W(t_0,t)(g,h) = W(0,t)(g)
\] 
exists, it is independent of $h(x)$, and for all reasonable initial
data $g(x)$ it gives rise to a 
well defined solution for all $t > 0$ emanating from the space-time
singularity at $t=0$.  Under reflection $t \to -t$ the Friedmann --
Robertson -- Walker metric gives a space-time metric for $t < 0$ with
a singular future at $t=0$, and the same solution formulae hold. We
thus have constructed solutions $u(t,x)$ of the wave equation in
Friedmann -- Robertson -- Walker space-times which exist for all
$-\infty < t < 0$ and $0 < t < +\infty$, where in conformally
regularized coordinates these solutions are continuous through the
singularity $t=0$ of space-time, taking on specified data 
$u(0,\cdot) = g(\cdot)$ at the singular time.  
\end{abstract}

\maketitle


\section{Introduction}

The family of Friedmann -- Robertson -- Walker metrics, Lorentzian metrics for
space-time which are spatially homogeneous, and which have a
singularity at $t=0$, have played a central r\^ole in general
relativity and cosmology~\cite{Hawking&Ellis73}. In particular they
provide the simplest case of space-times with a `big bang' 
singularity, and thus are important in the current interpretation of our
universe.  Lorentzian metrics with a singularity(without loss of
generality at $t=0$) have played a particular r\^ole in discussions of
the origins of space-time, having the striking feature of exhibiting
divergence of the curvature and the energy - momentum tensor, whether
in the past (`big bang') or at a future collapse of (a region of)
space-time. The linear wave equation in a background space-time metric 
describes the propagation of energy and matter fields in the linearized
regime, such as for example electromagnetic fields. It
is also the first model to consider when investigating the
propagation of linear fluctuations of the Einstein metric tensor
around this background. This article discusses a representation of the
wave propagator, and several of its consequences, for the initial
value problem for this wave equation, both for non-zero $t$ and for
the singular time $t=0$.  

In a short paper, Klainerman and Sarnak~\cite{Klainerman&Sarnak}
derived an explicit spherical means representation of the propagator
for the scalar wave equation in a Friedmann -- Robertson -- Walker
space-time in the case that the pressure and cosmological constant 
vanish (`dust' models), following ideas in F.~John~\cite{John,John2}. 
The Lorentzian line element corresponding to the classical Friedmann
-- Robertson -- Walker space-time metrics takes the form 
\begin{equation}\label{Eqn:LineElement}
   ds^2 = -dt^2 + S(t)^2d\sigma^2 ~,
\end{equation}
where $d\sigma^2$ is the line element for each underlying spatially
homogeneous time slice $\{(t,x) \, : \, t = t_0 \}$. In the cases we
consider these are space-like hypersurfaces corresponding to 
Euclidian space ${\mathbb R}^3$ in the case of curvature $K = 0$, and
to hyperbolic space ${\mathbb H}^3$ in the case of curvature $K = -1$. 
Under the time change 
\[
   \frac{dt}{d\tau} = S(t) ~, \qquad t(0) = 0 ~,
\]
the line element \eqref{Eqn:LineElement} is transformed to the form
\begin{equation}
   ds^2 = S(\tau)^2\bigl(-d\tau^2 + d\sigma^2 \bigr) ~,
\end{equation}
which is conformal to the half space ${\mathbb R}_+ \times {\mathbb
  R}^3 = \{ (t,x) : \tau > 0\}$ endowed with the
Minkowski metric. This conformal correspondence is of course singular
at $\tau = 0$ (which is the image of $t=0$), at which time the metric
interpretation is that all of space is contracted to a
point. Expressed in the transformed time variables $\tau$, the scale
factor $S(\tau)$ is given explicitly in~\cite{Hawking&Ellis73} 
\begin{eqnarray}\label{Eqn:ScaleFactor}
  && S(\tau) = \tau^2 ~, \qquad \hbox{when} \quad K = 0 ~, \\
  && S(\tau) = \cosh(\tau) - 1 ~, \qquad \hbox{when} \quad K = -1 ~.
   \nonumber
\end{eqnarray}

The wave propagator $W(\tau_0,\tau)(g,h)$ is the solution operator for
the wave equation 
\begin{equation}\label{Eqn:WaveEquation1}
   \Box u = 0 ~, \qquad u(\tau_0,x) = g(x) ~, \quad \partial_\tau u(\tau_0,x) =
   h(x) ~,   
\end{equation}
where the D'Alembertian operator in Friedmann -- Robertson -- Walker
space-times is given by 
\begin{equation}\label{Eqn:DAlembertian}
   \Box u = - \frac{1}{S^2} \partial_\tau^2 u - \frac{2\dot
     S}{S^3}\partial_\tau u + \frac{1}{S^2}\Delta_\sigma u ~.  
\end{equation}
Here $\Delta_\sigma$ is the Laplace -- Beltrami operator
corresponding to the Riemannian metric on the time slices $\{\tau =
\tau_0 > 0 \}$, in particular for the metrics in the two cases $K = 0$
and $K = -1$. Namely, given data $(g(x), h(x))$ in some appropriate class of
functions or distributions, the wave propagator $W(\tau_0,\tau_1)(g,h)$ is
defined to be the solution operator 
\[
    W{(\tau_0,\tau_1)}(g,h) := (u(\tau_1,x),\partial_\tau u(\tau_1,x)) ~,
    \qquad \tau_0 > 0 ~, \quad \tau_1 > 0 ~,
\]
where of course $u(\tau,x)$ is the solution to \eqref{Eqn:WaveEquation1}.

The object of this article is to draw several conclusions from the
spherical means expression for the solution operator for
\eqref{Eqn:WaveEquation1} given in reference \cite{Klainerman&Sarnak}. 
Using an explicit form of the propagator, a modification of that given
in~\cite{Klainerman&Sarnak}, we make three observations about solutions 
to the wave  equation in a Friedmann -- Robertson -- Walker background
metric. Firstly, we derive decay estimates for solutions in transformed
time $\tau \gg 0$ (and by consequence in physical time $t \gg 0$), given
compactly supported initial data $(g(x), h(x))$ posed on the Cauchy
surface $\{ \tau = \tau_0 > 0 \}$, using a remark of
F.~John~\cite{John}. In short, the result
is the following: for the case $K=0$ solutions of the wave equation
satisfy the estimate
\[
    |(u(t,\cdot),\partial_tu(t,\cdot)|_{L^\infty({\mathbb R}^3)} \leq \BigOh{t^{-1}}
\]
for large time $t$, a rate identical to that of four dimensional
Minkowski space-time. In the case of negatve curvature $K=-1$,
solutions obey a faster decay rate, namely
\[
  |(u(t,\cdot),\partial_tu(t,\cdot)|_{L^\infty({\mathbb H}^3)} \leq
  \BigOh{t^{-2}} ~.
\]  
Our second main result is that, while solutions satisfy the general
Huygen's principle of finite propagation speed, the sharp Huygen's 
principle is not valid, either for cases $K = 0$ or $K = -1$. This is
in contrast to the case of the Minkowski metric on 
${\mathbb R}^{1+3}$. The consequence is that signals do not propagate
sharply on the light cone, but instead, after the passage of the wave
front they leave a fading residual within the interior of the light
cone. More precisely, the support of the kernel of the wave propagator
is the full interior of the light cone, with the light cone itself
being its singular support. It is reminiscent of wave propagation
phenomena in the Minkowski metric for even space dimensions. This
phenomenon has been recently observed in \cite{Yagdjian} (2013) for
the Klein -- Gordon wave propagator in the de Sitter space-time
metrics. Thirdly, a principle feature of 
Friedmann -- Robertson -- Walker space-times are that they 
are singular at times $t = \tau = 0$, a fact that is central in our
current picture of cosmology. However it should not be assumed from
this that the initial value problem for the wave equation is not
without meaning for $t_0 = 0$. Indeed we show that for fixed
$t_1 > 0$ the limit of the wave propagator 
\[
    \lim_{t_0 \to 0 \atop t_0 > 0} W(t_0,t_1)(g,h) := W(0,t_1)(g)
\] 
exists, and gives rise to well defined solutions with admissible
initial data being precisely half of the Hilbert space of standard
Cauchy data for the wave equation. Under time reflection $t \mapsto -t$  
the Friedmann -- Robertson -- Walker space-times are also solutions of
Einstein's equations with the same properties of spatial homogeneity,
and the above initial value problem can just as well be run backwards
in time from $t = 0$, thus providing a full class of solutions of the
wave equation which are global in space-time, both for positive and
negative time $t \in {\mathbb R}$, with specified initial data $u(0,x)
= g(x)$ at $t=0$. In conformally regularized coordinates about $t = 0$
these solution are continuous, indeed smooth, through the time slice
$t = 0$, leading to the interpretation that they pass information
continuously through the singularity in space-time from the past to
the future.

\section{Wave Equation for $K=0$}

\subsection{The Solution operator in Minkowski space-time}

For comparison purposes we recall the spherical means representation
of solutions $u(t,x)$ of the initial value problem for the wave
equation in Minkowski space-time ${\mathbb R}^{1+3}$, namely
\begin{equation*}
   \partial_t^2 u - \Delta u = 0 ~, \qquad u(t_0,x) = g(x) ~, \quad
   \partial_t u(t_0,x) = h(x) ~.
\end{equation*}
Define the spherical means operator to be
\begin{equation}\label{Eqn:SphericalMean}
   M_{f}(r,x) := \frac{1}{4\pi
     r^{2}}\int_{S_r(x)} f(y) \, dS_{r}(y) ~.
\end{equation}
Then $rM_{u(t,\cdot)}(r,\cdot)$ satisfies the wave equation in one
space dimension. From this we deduce the spherical means
representation of the solution (also known as Kirchhoff's formula),
given by 
\begin{eqnarray}\label{Eqn:Kirchhoff}
    && u(t,x) = \partial_t((t-t_0)M_{g}(t-t_0,x)) + (t-t_0)M_{h}(t-t_0,x)  \nonumber  \\
      && \quad = \frac{1}{4\pi (t-t_0)} \int_{S_{t-t_0}(x)} \nabla g(y) \cdot
    \frac{y-x}{|y-x|} \ dS_{t-t_0}(y) 
  + \frac{1}{4\pi (t-t_0)^2} \int_{S_{t-t_0}(x)} g(y) \ dS_{t-t_0}(y) \\
  & & \qquad + \frac{1}{4\pi (t-t_0)} \int_{S_{t-t_0}(x)} h(y) \ dS_{t-t_0}(y) ~.
  \nonumber
\end{eqnarray}
The finite propagation speed of solutions (Huygen's principle),
the sharp Huygen's principle, and the time decay of solutions with
compactly supported initial data follow immediately from this
expression. Indeed, the solution at point $(t,x)$ is influenced only
by initial data $(g(x), h(x))$ at time $t=t_0$ at points on the sphere
$S_{t-t_0}(x)$, and in particular the integral kernel of the solution
operator at $(t,x)$ given by the spherical means formula \eqref{Eqn:Kirchhoff} 
is zero off of the lightcone $\{(t',x')\ : \ |t-t'|=|x-x'| \}$. 
A derivation and discussion of this expression of the solution to the
wave equation can be found in~\cite{John}\cite{John2}.

\subsection{The Solution operator in Friedmann -- Robertson -- Walker space-time}

For $K=0$ the scale factor \eqref{Eqn:ScaleFactor} takes the form 
$S(\tau)=\tau^{2}$,  thus the wave equation \eqref{Eqn:DAlembertian}
in the Friedmann -- Robertson -- Walker space-time takes the form: 
\begin{equation} 
\begin{cases}
    \partial_\tau^2 u + \frac{4}{\tau}\partial_\tau u - \Delta u = 0 &
   0 < \tau, \tau_{0} ~; \ x \in\mathbb{R}^{3}  \\
   u(\tau_{0},x) = g(x)  \\
   u_{\tau}(\tau_{0},x) = h(x)
\end{cases}\label{eq:1}
\end{equation}

Assume that $g\in C^{2}(\mathbb{R}^{3})$ and $h\in C^{1}(\mathbb{R}^{3})$,
and impose the condition that $\mathrm{supp}(g,h)\subseteq{B}_{R}(0)$
for some $R>0$, where ${B}_{R}(0)=\{x\in\mathbb{R}^{3}:\ |x|\leq R\}$, 
$|x|$ being the usual Euclidian distance. Equation \eqref{eq:1} has an
explicit expression for the general solution, indeed define a
transformed function 
\begin{equation}\label{Eqn:TransformedFunctionK=0}
   v(\tau,x) = \frac{1}{\tau}\partial_\tau(\tau^3 u) ~,
\end{equation}
which has an inverse expression for $u(\tau,x)$
\begin{equation}
   \tau^3 u(\tau,x) = \int_0^{\tau - \tau_0} (
      r + \tau_0) v(r + \tau_0, x) \, dr + \tau_0^3 g(x) ~. 
\end{equation}
The new function $v(\tau,x)$ satisfies the following wave equation
\begin{equation}\label{Eqn:WaveEquation2}
   \partial_\tau^2 v = \Delta v \qquad \tau, \tau_{0} > 0,\  x \in \mathbb{R}^{3}  
\end{equation}
as can be seen by a comparison with the ultrahyperbolic wave equation with 
five dimensional time variable, with solutions that are radial in
time. The initial data for \eqref{Eqn:WaveEquation2} is defined through
the transformation \eqref{Eqn:TransformedFunctionK=0};
\begin{equation}\label{Eqn:DataK=0}
  \begin{cases}
    v(\tau_{0},x) = 3\tau_{0}g(x) + \tau_{0}^{2}h(x) := \phi(x)  \\
    \partial_\tau v(\tau_{0},x) = 3g(x) + \tau_{0}^{2}\Delta g(x) 
      + \tau_{0}h(x) := \psi(x) ~.
  \end{cases}
\end{equation}
Equation \eqref{Eqn:WaveEquation2} is the wave equation in the Minkowski 
metric on the domain $\{(\tau, x) : \tau > 0, x \in {\mathbb R}^3\}$, whose  
solution operator is expressed by the spherical means formula~\eqref{Eqn:Kirchhoff}
\begin{equation}
  v(\tau,x) = \partial_{\tau}\left((\tau-\tau_{0})M_{\phi}(\tau-\tau_{0},x)\right)
    + (\tau-\tau_{0})M_{\psi}(\tau-\tau_{0},x) ~,
\end{equation}
where $M_{f}(r,x)$ is the spherical mean of
the function $f(x)$, (respectively, $\phi$ and $\psi$)~\eqref{Eqn:SphericalMean}.
Substituting the expression for $v(\tau,x)$ back into the expression
for $u(\tau,x)$ yields the following solution formula; 
\begin{align}
   \tau^{3}u(\tau,x) & = \int_{0}^{\tau-\tau_{0}} (r+\tau_{0})
      \partial_{r}\left(rM_{\phi}(r,x)\right)\, dr 
    + \int_{0}^{\tau-\tau_{0}} (r + \tau_{0})rM_{\psi}(r,x)\, dr  
    + \tau_{0}^{3}g(x)
\label{eq:2} \\
 & = \tau(\tau-\tau_{0})M_{\phi}(\tau-\tau_{0},x) 
   - \int_{0}^{\tau-\tau_{0}}rM_{\phi}(r,x)\, dr   \nonumber \\
 & \qquad + \int_{0}^{\tau-\tau_{0}}(r+\tau_{0})rM_{\psi}(r,x)dr 
   + \tau_{0}^{3}g(x) ~, \nonumber 
\end{align}
To present the formula in more useful form, consider the
initial data in separate cases: (1) $(g(x), h(x) = 0)$ and 
(2)~$(g(x) = 0, h(x))$. 

In case (1), using the definitions for $(\phi,\psi)$ given in
\eqref{Eqn:DataK=0}, the expression \eqref{eq:2} gives the formula  
\begin{eqnarray*}
   u(\tau,x) & = & \frac{1}{\tau^3} \biggl(\int_0^{\tau-\tau_0} 
      (r + \tau_0)\partial_r\bigl(\frac{1}{4\pi r}\int_{S_r(x)}
      3\tau_0 g(y) \ dS_r(y) \bigr) \ dr \\ 
   && + \int_0^{\tau-\tau_0} (r+\tau_0)r \bigl( \frac{1}{4\pi r^2} 
      \int_{S_r(x)} 3g(y) + \tau_0^2\Delta g(y) \ dS_r(y) \bigr) \ dr 
      + \tau_0^3 g(x) \biggr) \\ 
   & = & \frac{1}{\tau^3} \biggl(\int_0^{\tau-\tau_0} 
      \partial_r\bigl( (r + \tau_0)\frac{1}{4\pi r}
      \int_{S_r(x)} 3\tau_0 g(y) \ dS_r(y) \bigr) \ dr \\
   && + \int_0^{\tau-\tau_0} \frac{1}{4\pi} 
      \int_{S_r(x)} 3 g(y) \ dS_r(y)\ dr \\
   && + \int_0^{\tau-\tau_0} \frac{1}{4\pi} 
      \int_{S_r(x)} (\tau_0^2 + \frac{\tau_0^3}{r} )
      \Delta g(y) \ dS_r(y) \ dr + \tau_0^3 g(x) \biggr) ~.
\end{eqnarray*}
Applying the divergence theorem and using the fact that the
fundamental solution for the Laplacian is given by
$\frac{-1}{4\pi}\frac{1}{r}$, this gives the following expression for
the wave propagator: 
\begin{eqnarray}\label{Eqn:Expression-g-K=0}
   u(\tau,x) & = & \frac{1}{\tau^3} \biggl( \frac{1}{4\pi} 
     \frac{\tau^3 - (\tau - \tau_0)^3}{(\tau - \tau_0)^2} 
     \int_{S_{\tau - \tau_0}} g(y) \ dS_{\tau-\tau_0}(y)  \nonumber
     \\ 
     && + \frac{1}{4\pi} \frac{\tau_0^2\tau}{(\tau - \tau_0)}  
     \int_{S_{\tau - \tau_0}} \nabla g(y) \cdot \frac{y-x}{|y-x|}
     \ dS_{\tau-\tau_0}(y) \\ 
     && + \frac{3}{4\pi} \int_0^{\tau-\tau_0}\int_{S_r(x)} g(y)
     \ dS_{r}(y) \ dr  \biggr) ~.   \nonumber
\end{eqnarray}

In case (2), using \eqref{Eqn:DataK=0} the expression \eqref{eq:2}
gives rise to  
\begin{eqnarray}\label{Eqn:Expression-h-K=0}
   u(\tau,x) & = & \frac{1}{\tau^3} \biggl( \int_0^{\tau-\tau_0} 
       (r + \tau_0) \partial_r\bigl( \frac{1}{4\pi r} 
       \int_{S_r(x)} \tau_0^2 h(y) \ dS_r(y) \bigr) \ dr  \nonumber \\
    && + \int_0^{\tau-\tau_0} (r + \tau_0) \frac{1}{4\pi r} 
         \int_{S_r(x)} \tau_0 h(y) \ dS_r(y) \ dr  \biggr)  \\
    & = & \frac{1}{\tau^3} \biggl(
         \frac{\tau\tau_0^2}{4\pi(\tau-\tau_0)}  
         \int_{S_{\tau-\tau_0}(x)} h(y) \ dS_{\tau-\tau_0}(y) 
       + \frac{\tau_0}{4\pi} \int_0^{\tau-\tau_0}\int_{S_r(x)} h(y)
     \ dS_{r}(y) \ dr \biggr) ~.  \nonumber
\end{eqnarray}
The sum of the expressions \eqref{Eqn:Expression-g-K=0} and
\eqref{Eqn:Expression-h-K=0} gives the solution formula for the
wave propagator in the case of general initial data. In these
expressions the integral densities for surfaces and volumes are not
given with respect to the background Lorentzian metric $g$ restricted
to the time slice $\{(\tau,x) : \tau = \tau_0 \}$, which we denote 
$g_{\tau_0}$. This is implemented using the scale factor 
$S(\tau_0) = \tau_0^2$ and the substitutions 
$dS_{\tau-\tau_0} =
   S^{-2}(\tau_0)\bigl(S^2(\tau_0)dS_{\tau-\tau_0} \bigr) 
   := \tau_0^{-4} dS_{\tau-\tau_0}(g_{\tau_0})$, and 
$dS_{r}dr = S^{-3}(\tau_0)\bigl( S^3(\tau_0)dS_{r}dr \bigr) 
   := \tau_0^{-6} dV(g_{\tau_0})$. 
Additionally, the time variable $\tau$ is not the same as $t$ of the  
Friedmann -- Robertson -- Walker metric; it may be recovered
through the time change $t=\frac{\tau^{3}}{3}$. 

From the above expression for $u$, which contains integrals over the
interior of the backward light cone, one can see that for any given
$(\tau,x)$, $\tau > \tau_0$, the domain of dependence is ${B}_{\tau-\tau_{0}}(x)$. 
It then follows that the solution is identically zero outside of the 
union of interiors of all the future light cones emanating from
${B}_{R}(0)$, namely
 ${U_R}:=\{(\tau,x):\ |x|\leq R+(\tau-\tau_{0}),\ \tau>\tau_{0}\}$,
that is, $\mathrm{supp}(u) \subseteq {U_R}$, the statement of finite
propagation speed. In addition, consider the union of all future light
cones 
\[
   {V_R}=\underset{y\in B_{R}(0)}{\bigcup}\{(\tau,x):\ 
   |x-y| = \tau-\tau_{0},\ \tau>\tau_{0}\} ~.
\]
Based on the representation \eqref{Eqn:Expression-g-K=0} and
\eqref{Eqn:Expression-h-K=0} of the wave propagator, the solution
$u(\tau,x)$ is generally nonzero in $U_R \backslash V_R$, that is,
inside the envelope of light cones over the support of the initial
data, and it is given there by the expression
\begin{equation*}
    u(\tau,x) = \frac{1}{\tau^3} 
     \biggl( \frac{3}{4\pi} \int_0^{\tau-\tau_0}\int_{S_r(x)} g(y)
     \ dS_{\tau-\tau_0}(y) \ dr 
    + \frac{\tau_0}{4\pi} \int_0^{\tau-\tau_0}\int_{S_r(x)} h(y)
     \ dS_{\tau-\tau_0}(y) \ dr \biggr) ~,
\end{equation*}
valid for $(\tau,x) \in U_R\backslash V_R$.
One can see from these observations that for any given $x$, after
a given time (and in particular for $\tau>|x|+\tau_{0}+R$) the solution
at that point will generally persist indefinitely, spatially 
constant with value related to the the average value of the initial
data, however with asymptotically diminishing magnitude in time. Thus
the sharp Huygen's principle does not hold for solutions 
of the wave equation in this space-time, a fact which is in contrast
with the ordinary three dimensional wave equation, as is discussed in
pages 130-131 of \cite{John2}. Of course there is a similar statement
in the case  $0 < \tau < \tau_0$.

\subsection{Rate of Decay}

The explicit expression \eqref{eq:2} for the solution operator is
useful for estimates on the rate of decay of solutions as $\tau \to
\infty$, which we quantify in the following statement.

\begin{Thm}
Suppose $g\in C^{1}(\mathbb{R}^{3})$, $h\in C^{0}(\mathbb{R}^{3})$,
$\mathrm{supp}(g,h)\subset{B}_{R}(0)$. Then the solution
to (\ref{eq:1}) decays to zero at rate of $\mathcal{O}(\tau^{-3})$
uniformly throughout ${U}$ as $\tau$ tends to infinity.
\end{Thm}
Similar estimates of the decay rate hold for $\partial_\tau u(\tau,x)$,   
and therefore for the wave propagator $W(\tau_0,\tau)(g,h)$.   

\textbf{Proof.}
We consider initial data $g\in C^{1}(\mathbb{R}^{3})$, 
$h\in C^{0}(\mathbb{R}^{3})$, with
$\mathrm{supp}(g,h)\subseteq{B}_{R}(0)$.
Define constants 
\[
   C_{g} := \sup_{x\in\mathbb{R}^{3}}|(g(x), \nabla g(x))| =
   |g(x)|_{C^1(B_R)} , \ 
   C_{h} := \sup_{x\in\mathbb{R}^{3}}|h(x)| 
   = |h(x)|_{C^0(B_R)} ~,
\]
we show that
\[
   |u(\tau,x)| \leq \frac{ C_R}{\tau^3}(C_{g} + C_{h}) ~.
\]
To prove this, taking as a sample calculation we examine the decay
rate of the second term in the second equality of 
\eqref{Eqn:Expression-g-K=0}. The calculations for the remaining terms
follows similarly. 
\begin{align*}
  \frac{1}{\tau^{3}} 
      & \Bigl| \frac{1}{4\pi} \frac{\tau_0^2\tau}{(\tau - \tau_0)}  
        \int_{S_{\tau - \tau_0}} \nabla g(y) \cdot \frac{y-x}{|y-x|}
        \ dS_{\tau-\tau_0}(y) \Bigr| \\
  & \leq \frac{1}{\tau^{3}} \Bigl| \frac{\tau_0^2\tau}{(\tau - \tau_0)} \Bigr| 
       \frac{C_{g}}{4\pi}   \int_{S_{\tau-\tau_{0}}(x) \cap B_R(0)} \ dS_{\tau-\tau_{0}}(y)   
  \leq \frac{C_{g} R^2}{\tau^{3}} = \BigOh{\tau^{-3}} ~.
\end{align*}
Note that the term $|{\tau_0^2\tau}/{(\tau - \tau_0)}| \sim \tau_0^2$
for large $\tau$, and that  
$|\int_{S_{\tau-\tau_{0}}(x) \cap B_R(0)} \ dS_{\tau-\tau_{0}}(y)|$ is
bounded by $4\pi \min\{ (\tau - \tau_0)^2, R^2\}$ for geometrical reasons.
In the second line we have used that the support of the initial data
is compact, indeed $\mathrm{supp}(g)\subseteq B_{R}(0)$. 

A similar analysis handles the remaining terms, yielding
\begin{align*}
 & \frac{1}{\tau^{3}} \Bigl| \frac{1}{4\pi} 
     \frac{\tau^3 - (\tau - \tau_0)^3}{(\tau - \tau_0)^2} 
     \int_{S_{\tau - \tau_0}} g(y) \ dS_{\tau-\tau_0}(y) \Bigr|  
   = \BigOh{R^2\tau^{-3}}   \\
 & \frac{1}{\tau^{3}} \Bigl| \frac{3}{4\pi} \int_0^{\tau-\tau_0}\int_{S_r(x)} g(y)
     \ dS_{\tau-\tau_0}(y) \ dr \Bigr|
   = \BigOh{R^3\tau^{-3}}    \\ 
 & \frac{1}{\tau^{3}}  \Bigl| \frac{\tau\tau_0^2}{4\pi(\tau-\tau_0)}  
         \int_{S_{\tau-\tau_0}(x)} h(y) \ dS_{\tau-\tau_0}(y)  \Bigr|  = \BigOh{R^2\tau^{-3}} \\
 & \frac{1}{\tau^{3}}  \Bigl| \frac{\tau_0}{4\pi} \int_0^{\tau-\tau_0}\int_{S_r(x)} h(y)
     \ dS_{\tau-\tau_0}(y) \ dr \Bigr|  = \BigOh{R^3\tau^{-3}}
\end{align*}
for large $\tau \mapsto +\infty$. The result is that for $(\tau,x)\in
{U}$, for $\tau$ large, $u(\tau,x) = \BigOh{\tau^{-3}}$, and uniformly
so throughout ${U}$.  \hfil $\Box$

To recover a decay estimate in terms of our original time variable
$t$, use the fact that $t = (\tau^{3}/3)$, implying that 
$\tau=(3t)^{\frac{1}{3}}$. Therefore the decay of the solution of the 
wave equation in flat Friedmann - Robertson - Walker space-time is
$\BigOh{t^{-1})}$, which is identical to the rate of decay of
solutions for the wave equation for the Minkowski metric in three
dimensional space.

\section{Wave equation for $K=-1$}

\subsection{The Solution}

\noindent We now solve the wave equation in the Robertson-Walker space-time
for constant curvature $K=-1$. We know in this case the scaling factor
takes the form $S(\tau)=\cosh(\tau)-1$, gving rise to the following
system
\begin{align}\label{eq:4}
  && \partial_\tau^2 u + 2\coth(\frac{\tau}{2}) \partial_\tau u 
  - \Delta_\sigma u = 0 ~, \quad  \tau > 0 ~, \quad  x \in\mathbb{H}^{3}   \\
  &&  u(\tau_{0},x) = g(x) ~, \quad 
   \partial_\tau u(\tau_{0},x) = h(x)~, \qquad \tau_{0} > 0 ~.  \nonumber 
\end{align}
Assume similar support constraints to our initial data as in
the case for $K=0$, namely that $(g(x), h(x))$ are supported in the
geodesic ball ${B}_{R}(0)$. Then equation \eqref{eq:4} has an explicit 
geodesic spherical means expression for solutions, similar to the one
given in \eqref{eq:2}. Define a transformed function
\begin{equation}\label{Eqn:TransformedFunctionK=-1}
   v(\tau,x) = \frac{4}{\sinh{(\frac{\tau}{2})}} \partial_\tau 
     \bigl(\sinh^3(\frac{\tau}{2}) u(\tau,x) \bigr) ~, 
\end{equation}
in analogy with \eqref{Eqn:TransformedFunctionK=0}. The inverse of
this transformation is given by
\begin{equation}\label{Eqn:InverseTransformedFunctionK=-1}
   \sinh^3(\frac{\tau}{2}) u(\tau,x) = \int_0^{\tau - \tau_0}
   \frac{1}{4} \sinh(\frac{r + \tau_0}{2}) v(r + \tau_0,x) \, dr 
   + \sinh^3(\frac{\tau_0}{2}) u(\tau_0,x) ~. 
\end{equation}
Then $v(\tau,x)$ satisfies the equation
\begin{equation}\label{Eqn:WaveEqnK=-1}
  \partial_\tau^2 v = L v 
\end{equation}
with $L = \Delta_\sigma + 1$ and with Cauchy data given on the
hypersurface $\tau = \tau_0 > 0$ 
\begin{eqnarray}\label{Eqn:CauchyDataK=-1}
  && v(\tau_0,x) = 3 \sinh(\tau_0) g(x) + 4 \sinh^2(\frac{\tau_0}{2})
    h(x) := \phi(x)  \\
  && \partial_\tau v(\tau_0, x) = 3 \cosh(\tau_0) g(x) 
     + 4\sinh^2(\frac{\tau_0}{2}) \Delta_\sigma g + \sinh(\tau_0) h(x)
     := \psi(x)  ~.   \nonumber 
\end{eqnarray}
For \eqref{Eqn:WaveEqnK=-1} there is an explicit spherical means
formula for the solution, given in \cite{Lax&Philips}, which is the
hyperbolic analog of \eqref{Eqn:Kirchhoff}; namely
\begin{equation}\label{Eqn:WaveSolutionK=-1}
   v(\tau,x) = \sinh(\tau - \tau_0) M_\psi(\tau - \tau_0,x) +
   \partial_\tau \bigl( \sinh(\tau - \tau_0) M_\phi(\tau - \tau_0,x)\bigr) ~,
\end{equation}
where in the case $K = -1$ the geodesic spherical mean of a function
$f(x)$ is given by an integral over the geodesic sphere $S_r(x)$ of radius $r$
about $x$;
\[
    M_{f}(r,x) := \frac{1}{4\pi (\sinh(r))^2} \int_{S_r(x)} f(y) \,
    dS_r(y) ~,
\]
and where $dS_r(x)$ is the element of spherical surface area.  Using 
\eqref{Eqn:WaveSolutionK=-1} and the inversion formula 
\eqref{Eqn:InverseTransformedFunctionK=-1}, the solution operator can
be expressed as  
\begin{align}\label{Eqn:SphericalMeansWaveOPK=-1}
    \sinh^{3}(\frac{\tau}{2})u(\tau,x) &
  = \int_{0}^{\tau-\tau_{0}}\frac{1}{4} \sinh(\frac{r + \tau_{0}}{2})
     \partial_{r}\bigl( \sinh(r)M_{\phi}(r,x) \bigr) \, dr  \nonumber   
   \\ 
   &  \quad + \int_{0}^{\tau-\tau_{0}}\frac{1}{4} \sinh(\frac{r + \tau_{0}}{2})
     \sinh(r)M_{\psi}(r,x) \, dr  \\ 
   &  \quad + \sinh^{3}(\frac{\tau_{0}}{2})g(x)  ~.  \nonumber 
\end{align}
Using the definitions for $(\phi,\psi)$ given in \eqref{Eqn:CauchyDataK=-1}, 
this expresses the solution of the wave propagator for the hyperbolic
case in terms of spherical means over geodesic spheres in hyperbolic
geometry. A more transparent expression is obtained from substituting
the actual initial data. As in the Euclidian case, this is separated
into the two cases: (1) $(g(x), h(x)=0)$, and (2) $(g(x)=0, h(x))$. 

In case (1), the formula \eqref{Eqn:SphericalMeansWaveOPK=-1} gives
the expression 
\begin{eqnarray}\label{Eqn:computation1}
    u(\tau,x) & = & \frac{1}{\sinh^3(\tau/2)}\biggl( \int_0^{\tau-\tau_0} 
       \frac{3}{4} \sinh(\frac{r+\tau_0}{2})\sinh(r)\cosh(\tau_0) 
          M_g(r,x) \ dr  \nonumber \\
    && + \int_0^{\tau-\tau_0}  \frac{3}{4} \sinh(\frac{r+\tau_0}{2}) 
         \partial_r\Bigl( \sinh(r) \sinh(\tau_0) M_g(r,x) \Bigr) \ dr
         \\ 
    && + \int_0^{\tau-\tau_0} \sinh(\frac{r+\tau_0}{2}) \sinh(r) \sinh^2(\tau_0/2) 
         M_{\Delta_\sigma g}(r,x) \ dr + \sinh^3(\tau_0/2) g(x)
         \biggr)  \nonumber \\  
     & = & \frac{1}{\sinh^3(\tau/2)}\biggl( 
         \frac{3}{4} \sinh(\tau/2) \sinh(\tau-\tau_0) \sinh(\tau_0)
         M_g(\tau-\tau_0,x) \nonumber \\  
    && + \int_0^{\tau-\tau_0} \frac{3}{8}
         \Bigl(\sinh(\frac{r+\tau_0}{2})\cosh(\tau_0) +
         \sinh(\frac{r-\tau_0}{2}) \Bigr) \sinh(r) M_g(r,x) \ dr
         \nonumber  \\
   && + \int_0^{\tau-\tau_0} \sinh(\frac{r+\tau_0}{2}) \sinh(r) \sinh^2(\tau_0/2) 
         M_{\Delta_\sigma g}(r,x) \ dr + \sinh^3(\tau_0/2) g(x)
         \biggr) ~.  \nonumber 
\end{eqnarray}
The term involving the geodesic spherical mean of the Laplace --
Beltrami operator is treated by integrations by parts. Using the
fact that the fundamental solution is given by $\frac{-1}{4\pi \sinh(r)}$ 
and the definition of the geodesic spherical mean, we find
\begin{eqnarray*}
  && \int_0^{\tau-\tau_0} \sinh(\frac{r+\tau_0}{2}) \sinh(r) \sinh^2(\tau_0/2) 
         M_{\Delta_\sigma g}(r,x) \ dr \\
  && = \int_{B_{\tau-\tau_0}(x)}\sinh(\frac{r+\tau_0}{2})\sinh^2(\tau_0/2)  
      \frac{1}{4\pi\sinh(r)}\Bigl((\partial_r^2  + 2\coth(r)
      \partial_r) g \Bigr) \sinh^2(r) \ dS_1(\xi) \ dr   \\
  && = \frac{\sinh^2(\tau_0/2)\sinh(\tau/2)}{4\pi \sinh(\tau-\tau_0)}   
       \int_{S_{\tau-\tau_0}(x)} \partial_r g(y) \ dS_{\tau-\tau_0}(y)  \\
  && \quad +  
       \frac{\sinh^2(\tau_0/2)\bigl(\sinh(\tau/2)\cosh(\tau-\tau_0) 
     - \frac{1}{2}\cosh(\tau/2)\sinh(\tau-\tau_0)\bigr)}{4\pi\sinh^2(\tau-\tau_0)}
       \int_{S_{\tau-\tau_0}(x)} g(y) \ dS_{\tau-\tau_0}(y)  \\
  && \quad - \int_0^{\tau-\tau_0} \int_{S_{r}(x)} \frac{3}{4} \sinh^2(\tau_0/2)
       \sinh(\frac{r+\tau_0}{2})\frac{1}{4\pi\sinh(r)} g(y) \ dS_r(y) \ dr 
     - \sinh^3(\tau_0/2) g(x) ~.
\end{eqnarray*} 
Using this in the expression \eqref{Eqn:computation1}, one finds that
\begin{eqnarray}\label{Eqn:WaveOpCase1}
   && u(\tau,x) = \frac{1}{\sinh^3(\tau/2)}\Biggl( 
       \frac{\sinh^2(\tau_0/2)\sinh(\tau/2)}{4\pi \sinh(\tau-\tau_0)}   
       \int_{S_{\tau-\tau_0}(x)} \partial_r g(y) \ dS_{\tau-\tau_0}(y)
       \nonumber \\
  && \quad  
   + \Bigl( \frac{1}{2}\sinh(\tau/2)\sinh(\tau_0/2)\bigl(3\sinh(\tau-\tau_0)\cosh(\tau_0/2) 
           + \cosh(\tau-\tau_0)\sinh(\tau_0/2) \bigr) \\
  && \quad\quad
   + \frac{1}{2}\sinh^3(\tau_0/2)\Bigr)
           \frac{1}{4\pi\sinh^2(\tau-\tau_0)}
           \int_{S_{\tau-\tau_0}}g(y) \ dS_{\tau-\tau_0}(y) \nonumber \\
  && \quad
   + \int_0^{\tau-\tau_0} \int_{S_{r}(x)} \frac{3}{8} \Bigl(
           \sinh(\frac{r+\tau_0}{2}) + \sinh(\frac{r-\tau_0}{2}) \Bigr) 
           \frac{1}{4\pi\sinh(r)} g(y)
           \ dS_r(y) \ dr \Biggr) ~.  \nonumber
\end{eqnarray}

In the case (2), the expression  \eqref{Eqn:SphericalMeansWaveOPK=-1}
is the following
\begin{eqnarray}\label{Eqn:WaveOpCase2}
    u(\tau,x) & = & \frac{1}{\sinh^3(\tau/2)} \biggl(  
    \int_0^{\tau-\tau_0} \sinh(\frac{r + \tau_0}{2}) 
      \partial_r\Bigl(\sinh(r) \sinh^2(\frac{\tau_0^2}{2}) M_h(r,x)
      \Bigr) \ dr  \nonumber \\
   && + \int_0^{\tau-\tau_0} \frac{1}{4} \sinh(\frac{r + \tau_0}{2}) 
       \sinh(r) \sinh(\tau_0) M_h(r,x) \ dr \biggr)  \nonumber\\
   & = & \frac{1}{\sinh^3(\tau/2)} \biggl( 
       \frac{\sinh(\tau/2))\sinh^2(\tau_0/2))}{4\pi\sinh(\tau-\tau_0)} 
       \int_{S_{\tau-\tau_0}(x)} h(y) \ dS_{\tau-\tau_0}(y)  \\   
   && + \int_0^{\tau-\tau_0} \int_{S_{r}(x)} \frac{1}{4}
      \Bigl(\cosh(\frac{r+\tau_0}{2}) - \cosh(\frac{r-\tau_0}{2}) \Bigr) 
     \frac{1}{4\pi\sinh(r)} h(y) \ dS_{r}(y) \ dr \biggr) ~. \nonumber  
\end{eqnarray}
In the expressions \eqref{Eqn:WaveOpCase1}\eqref{Eqn:WaveOpCase2} the
integral densities for surfaces and volumes are not 
given with respect to the background Lorentzian metric $g$ restricted
to the time slice $\{(\tau,x) : \tau = \tau_0 \}$; this is implemented
using the scale factor $S(\tau_0) = \cosh(\tau_0) - 1 = 2\sinh^2(\tau_0/2)$ 
and the substitutions 
$dS_{\tau-\tau_0} =
   S^{-2}(\tau_0)\bigl(S^2(\tau_0)dS_{\tau-\tau_0} \bigr) 
   := \frac{1}{4}\sinh^{-4}(\tau_0/2) dS_{\tau-\tau_0}(g_{\tau_0})$, and 
$dS_{r}dr = S^{-3}(\tau_0)\bigl( S^3(\tau_0)dS_{r}dr \bigr) 
   := \frac{1}{8}\sinh^{-6}(\tau_0/2) dV(g_{\tau_0})$. 
Additionally, the time variable $\tau$ is not the same as $t$ of the  
Friedmann -- Robertson -- Walker metric; it may be recovered by
inverting the time change $t = \sinh(\tau) - \tau$. 

Similar to the case for $K=0$, the domain of dependence of the solution
to the case $K = -1$ is the geodesic ball ${B}_{\tau-\tau_{0}}(0)$,
so that the support of the solution is restricted to
$\mathrm{supp}(u)= {U_R} := \{(\tau,x):\ dist_{\mathbb H}(x,0)
\leq R + \sinh(\tau-\tau_{0}),\ \tau > \tau_{0}\}$. Because of the final
terms of the RHS in both of the expressions
\eqref{Eqn:WaveOpCase1}\eqref{Eqn:WaveOpCase2}, it is evident that the
support of the kernel of the wave propagator is not confined to the
set $V_R := \cup_{y\in B_r(0)} \{ (\tau,x) : dist_{\mathbb H}(x,y)=\tau-\tau_0 \}$ 
of light-cones eminating from the initial data, rather it fills
the interior of the future light cone. Namely, the sharp Huygen's
principle does not hold in the negative curvature case, and in
particular, solutions of the wave equation in the hyperbolic case of
Friedmann - Robertson - Walker space-times do not experience sharp
propagation of signals.  Furthermore, and in contrast to the case of
$K=0$, the solution in the interior of the light cone $U_R\backslash V_R$ 
is not locally spatially constant, and is dictated by a kernel which
is dependent upon the geodesic radius $r$. Again, there is a similar
statement for the case $0 < \tau < \tau_0$. 

\subsection{Rate of Decay}
The explicit expression via spherical means over geodesic spheres
gives information about the decay of solutions for large times. This
is quantified in the following statement.   

\begin{Thm}
Suppose $g \in C^{1}(\mathbb{H}^{3})$, $h \in C^{0}(\mathbb{H}^{3})$,
with support $\mathrm{supp}(g,h) \subset {B}_{R}(0)$. Then the solution
to \eqref{eq:1} decays to zero at rate of $\BigOh{\sinh^2(R)e^{-{2}\tau}}$
uniformly throughout ${U}_R$ as $\tau$ tends to infinity.
\end{Thm}

Similar estimates of the decay rate hold for $\partial_\tau u(\tau,x)$,   
and therefore for the wave propagator $W(\tau_0,\tau)(g,h)$. Because
of the relation \eqref{Eqn:ScaleFactor} between conformal time $\tau$
and physical time $t$ in the hyperbolic case, the result is that
solutions have the decay rate for large $t$
\begin{equation}
   |W(t_0,t)(g,h)|_{L^\infty} \leq \BigOh{t^{-2}} ~.
\end{equation}

\textbf{Proof.}
We are assuming that $g\in C^{1}(\mathbb{H}^{3})$, $h\in C^{0}(\mathbb{H}^{3})$ and 
$\mathrm{supp}(g,h) \subset {B}_{R}(0)$. Define constants 
\[
   C_{g} := \sup_{x\in\mathbb{H}^{3}} \
   |(g(x), \nabla g(x))| \qquad 
   C_{h} := \sup_{x\in\mathbb{H}^{3}}|h(x)| ~.
\]
The result follows from estimates of the terms of the spherical means
expression \eqref{Eqn:WaveOpCase1}\eqref{Eqn:WaveOpCase2}. As a sample
calculation, examine the decay rate of the first term of the RHS of
\eqref{Eqn:WaveOpCase1}. The estimates for the remaining terms will
follow similarly. Using the compact support of $g$, the domain of
integration is $B_{\tau-\tau_{0}}(x) \bigcap B_{R}(0)$.  
\begin{eqnarray*}
  && \biggl| 
       \frac{\sinh^2(\tau_0/2)\sinh(\tau/2)}{4\pi
         \sinh^3(\tau/2)\sinh(\tau-\tau_0)}    
       \int_{S_{\tau-\tau_0}(x)} \partial_r g(y) \ dS_{\tau-\tau_0}(y)
       \biggr|  \\
  && \quad \leq \frac{\sinh^2(\tau_0/2)}{4\pi
         \sinh^2(\tau/2)\sinh(\tau-\tau_0)} 
         \biggl| \int_{S_{\tau-\tau_0}(x)\cap B_R(0)} 
            \nabla g \cdot N \ dS_{\tau-\tau_0}(y)\biggr|  ~,
\end{eqnarray*}
where $N$ is the exterior unit normal to the geodesic sphere $S_R(0)$.
By hypotheses $|\nabla \cdot g(y)| \leq C_g$. Then, for geometrical reasons,
the integral term is bounded by 
$C_g 4\pi \min\{\sinh^2(\tau-\tau_0),\sinh^2(R))\}$. Finally, the
first factor is bounded for large $\tau$ by $e^{-2\tau}$. A similar
analysis applies to the other four terms of \eqref{Eqn:WaveOpCase1}
and \eqref{Eqn:WaveOpCase2}. For $(\tau,x) \in U_R$ and for $\tau$
large, the conclusion is that $|u(\tau,x)| \leq \BigOh{e^{-2\tau}}$. 
Of course for $(\tau,x) \not\in U_R$ the solution is $u(\tau,x)=0$. 
\hfil $\Box$ 

Returning to physical variables, $t = \sinh(\tau)-\tau$, however
unlike the case for $K=0$, there is not a clean expression for the 
conformal time $\tau$ in terms of physical time $t$. The asymptotics
of this expression are such that, for any $a>0$ there is $t_{\star}>0$
such that for $t>t_{\star}$, $\ln(t) - a < \tau(t) < \ln(t) + a$. 
Thus $e^{-2\tau(t)} \sim e^{-{2}\ln (t)} = t^{-{2}}$, giving the decay
of the solution of the wave equation in Friedmann - Robertson - Walker 
space-time, expressed in the physical time variables $t$, as being 
$\BigOh{t^{-{2}}}$.

\section{The initial value problem at the singular time $\tau_{0}=0$}
\label{Sect:IVPat0}

Throughout the discussion above we have retained the condition that
$\tau, \tau_0 > 0$, due to the space-time metric singularity at
$\tau=0$ (equivalently at $t=0$). However, given the explicit nature
of the wave propagator $W(\tau_0,\tau)(g,h)$, we are able to consider
the limit  
\[
   \lim_{\tau_0 \to 0+} W(\tau_0,\tau_1)(g,h)
\]
where the initial time $\tau_0$ is taken to zero while leaving time
$\tau_1$ fixed. Since in both cases $K=0$ and $K=-1$ the 
transformed time variable behaves asymptotically as $t \sim \tau^3/3$ for
$\tau \to 0$ in a neighborhood of $t = 0 = \tau$, it suffices to work
in the transformed time $\tau$.  

\subsection{\bf Case $K=0$}
For zero curvature $K=0$ and $\tau_{0}>0$ the solution to the Cauchy
problem takes the form 
\[
    u(\tau,x) = \frac{1}{\tau^{3}}\int_{0}^{\tau-\tau_{0}} (r + \tau_{0})
    \partial_{r}\left( rM_{\phi}(r,x) \right)\, dr
   + \frac{1}{\tau^{3}} \int_{0}^{\tau-\tau_{0}} (r + \tau_{0}) 
     rM_{\psi}(r,x)\, dr + \frac{1}{\tau^{3}} \tau_{0}^{3}g(x) ~.
\]
Using the definition in \eqref{Eqn:WaveEquation2} for
$(\phi(x),\psi(x))$, recalling that they themselves depend upon
$\tau_0$, and taking the limit $\tau_{0}\rightarrow 0$ of the
resulting formula yields the following limiting expression for the
solution: 
\begin{equation}\label{Eqn:ReducedIVP}
   u(\tau,x) = \frac{1}{\tau^{3}}\int_{0}^{\tau}3r^{2}M_{g}(r,x) \, dr
\end{equation}
Remark that the limit as $\tau_0$ of this solution has no dependence
on the initial data $h(x)$. 

\begin{Thm}
For $(g,h) \in C^1({\mathbb H}^3)\times  C^0({\mathbb H}^3)$ the limit
of the wave propagator exists,  
\[
   \lim_{\tau_0 \to 0+} W(0,\tau)(g,h) 
   = \bigl( u(\tau,x),\partial_\tau u(\tau,x) \bigr) ~, 
\]
it is independent of $h(x)$, and satisfies
\[
   \lim_{\tau \to 0+} u(\tau,x) = g(x) ~, \quad  
   \lim_{\tau \to 0+} \partial_\tau u(\tau,x) = 0 ~. 
\]
Thus the expression \eqref{Eqn:ReducedIVP} gives a solution to the
wave equation over the full half-line $\tau \in (0,+\infty)$, with
initial data $(u(0,x), \partial_\tau u(0,x)) = (g(x),0)$ given at the
singular time $\tau = 0$. 
\end{Thm}

\textbf{Proof.}
From the expression \eqref{Eqn:ReducedIVP} and l'H\^opital's rule, one
finds that 
\[
   \lim_{\tau \to 0+} u(\tau,x) = \lim_{\tau \to 0+} M_g(\tau,x) = g(x)
\]
for continuous initial data $g(x)$. Furthermore, one verifies that
expression \eqref{Eqn:ReducedIVP} satisfies 
\[
   \lim_{\tau\rightarrow 0}\partial_{\tau} u(\tau,x) = 0 ~.
\]
Namely, after differentiation, the expression for $\partial_{\tau} u(\tau,x)$ 
is explicitly  
\begin{equation*}
  \partial_{\tau} u(\tau,x) = \frac{1}{\tau^3} 
     \Bigl(\frac{-3}{\tau} \int_0^\tau 3 r^2 M_g(r,x)\, dr + 3\tau^2
     M_g(\tau,x) \Bigr) ~, 
\end{equation*}
for which, using l'H\^opital's rule,  one has
\[
  \lim_{\tau\rightarrow 0}\partial_{\tau} u(\tau,x) = 
  \lim_{\tau\rightarrow 0} \frac{3}{4}\partial_\tau M_g(\tau,x) ~, 
\]
and this vanishes as $\tau$ tends to zero since $M_g(\tau,x)$ is even
in $\tau$. \hfil $\Box$  


It is clear from this calculation that the full Cauchy problem is not
well posed for $\tau_{0}=0$. In general, solutions of the Cauchy
problem posed at $\tau_0 > 0$ and propagated to times $0 < \tau < \tau_0$ 
will become singular as $\tau \to 0+$. But there remains a full function
space of initial data, depending upon one scalar function $g(x)$, for
which the solution exists for the initial value problem consisting of
data $(g(x),0)$ posed at $\tau_0 = 0$, and propagated to arbitrary
future (or past) times $\tau$. 

\subsection{$K=-1$}
There is a similar calculation of the limit as $\tau_0 \to 0$ for the
case of constant curvature $K=-1$. For $\tau_{0}>0$ the solution to
the Cauchy problem takes the form 
\begin{align*}
   u(x,\tau) & = \frac{1}{4\sinh^{3}(\frac{\tau}{2})}\int_{0}^{\tau-\tau_{0}}
       \sinh(\frac{r + \tau_{0}}{2}) \partial_{r}(\sinh(r)M_{\phi}(r,x))\, dr  \\
    & \quad + \frac{1}{4\sinh^{3}(\frac{\tau}{2})}\int_{0}^{\tau-\tau_{0}}
       \sinh(\frac{r + \tau_{0}}{2}) \sinh(r)M_{\psi}(r,x)\, dr 
    + \frac{\sinh^{3}(\frac{\tau_{0}}{2})}{\sinh^{3}(\frac{\tau}{2})}g(x)
\end{align*}
Again recall that the Cauchy data $(\phi(x),\psi(x))$ given in
\eqref{Eqn:CauchyDataK=-1} are defined in terms of $\tau_0$. Taking
the limit $\tau_{0}\rightarrow0$ of this expression, after a similar
short calculation, yields the following expression for a solution:
\begin{equation}\label{Eqn:SolutionK=-1t=0}
   u(\tau,x) = \frac{1}{4\sinh^{3}(\frac{\tau}{2})} 
      \int_{0}^{\tau} 3\sinh(\frac{r}{2})\sinh(r)M_{g}(r,x) \, dr ~.
\end{equation}
Thus, similarly to the case $K=0$ the Cauchy problem for $\tau_{0} = 0$ is
ill-posed for arbitrary initial data. However for the particular
initial data $(g(x), 0)$ there is a well defined solution that 
initiates from $\tau_0 = 0$, given by the expression
\eqref{Eqn:SolutionK=-1t=0}. Its time derivative is well behaved in
the limit, indeed 
\[
    \lim_{\tau\rightarrow 0} u(\tau,x) = g(x) ~, \quad
    \lim_{\tau\rightarrow 0} \partial_{\tau} u(\tau,x) = 0 ~,
\]
as is shown by short calculations similar to those of the previous
section. This gives a meaning to the wave propagator when applied to  
data $(g(x), 0)$ posed at $\tau = 0$, resulting in a well
defined solution for $\tau > 0$ emanating from the singularity of
space-time at $\tau_0 = t_0 = 0$. Under reflection $\tau \to -\tau$
this is also a solution to the wave equations \eqref{eq:1},
respectively \eqref{eq:4} for $\tau < 0$, for which the evolution is
continuous across the space-time singularity at $\tau=0$. 

\section{Perspectives}\label{Sect:Perspectives}

The analysis in this article of the wave propagator in a Friedmann --
Robertson -- Walker background space-time metric raises a number of
basic questions, having to do with the propagation of signals in an
expanding universe, as well as having to do with the passage of
information through a space-time singularity from the past to the
future. The most basic questions are as follows. 

\hskip 20pt
The result of Section~\ref{Sect:IVPat0} is that certain information
propagated by solutions of the wave equation can be transmitted
continuously from the past to the future of a space-time singularity
of the character of the singularity in a Friedmann -- Robertson -- Walker 
space-time. As the wave equation \eqref{Eqn:WaveEquation1} represents
the linearized theory of the evolution of many physical quantities,
this class of solutions is analogous to the theory of a linear stable
manifold for the wave equation at the singularity of the Friedmann --
Robertson -- Walker space-times. It would be a fundamental question
whether nontrivial families of solutions of nonlinear equations exist 
which pass in a similar way through a Friedmann -- Robertson -- Walker-like
space-time singularity, in a way which transmits information from past
to future. In particular, one should ask whether there exist families
of solutions to the full Einstein equations which behave similarly,
and which are regularized under an appropriate singular conformal
transformation. these would carry a nontrivial past  space-time into a
nontrivial future. Following the analogy with stable manifold theory,
the indication is that a large class of such solutions could exist. 

\hskip 20pt
A second question has to do with the measurement of large scale
distances based on decay properties of the intensity of supernovae. In
large sky surveys using type $Ia$ supernovae as standard candles, two
basic pieces of information are compared to high precision: one is
redshift, giving a precise measurement of relative velocity of the
supernova away from the observer. The second is a measurement of the
intensity of the supernova, from which we deduce distance from the
observer. Given data from the second, observations show that the red
shift, and therefore the relative velocity of distant objects, is not
consistent with the Hubble hypothesis of a uniformly expanding
universe, as this velocity is slightly and unexpectedly higher for
more distant objects. This is the phenomenon of apparent acceleration
of the universe, which gives rise to the idea of the possibility of a
nonzero cosmological constant, among other theories. With the
precision of the description of solutions to the wave equation given
in this article, and the decay rates that follow from it, it would
merit another look at this study. Namely, a reassessment in particular
of the transformation between the measurement of supernova intensity
and the distance from the observer. To apply our detailed results on
the wave propagator, we would make the assumption that
on a sufficiently large scale the space-time 
metric is close to the Friedmann -- Robertson -- Walker metric. The
decay rates in fact involve the parameter of the curvature $K$, where
in this article we have described the two cases $K=0$ and $K=-1$. Of
course an expression  from the same derivation for the wave propagator
for any $0 \geq K > -\infty$ is available after a scaling. It is
conceivable, assuming a Friedmann -- Robertson -- Walker background
and therefore a fixed Hubble expansion rate, that the result could be
an improved fit of the data without the apparent acceleration, and where
the best fit for the distance data would also give an observed value
for the space-time curvature $K$. 
\looseness=-100

\hskip 20pt
A third question has to do with the lack of the sharp Huygens
principle with a Friedmann -- Robertson -- Walker background metric,
and whether it can be seen in observations. This article has only
addressed the initial value problem for the wave equation, but by the
Duhamel principle the inhomogeneously forced problem can also be
expressed in terms of the wave propagator. This would be what is used
to describe light from a steady distant source impinging over a long
period of time on our location on Earth. Given the lack of a sharp
Huygens principle, a wave front which has passed a location will
leave a small residual in the form of a decaying trace. We propose
that an observer could conceivably see in this in an effect 
on the red shift of spectral lines; the trace left by the passage in
the past of a wave front, if still detectible, would be of a slightly
smaller red shift, in the form of a slight blue downshifted trace or 
shadow, due to the slightly smaller velocity of recession of the
source in the distant past when the wave front was emitted.





\frenchspacing
\bibliographystyle{plain}

\end{document}